\documentstyle[12pt,epsf]{article}
\begin{document}
\input epsf.tex

\newcommand{\nc}{\newcommand}
\nc{\beq}{\begin{equation}}
\nc{\eeq}{\end{equation}}
\nc{\beqa}{\begin{eqnarray}}
\nc{\eeqa}{\end{eqnarray}}
\nc{\lra}{\leftrightarrow}
\nc{\ra}{\rightarrow}
\nc{\sss}{\scriptscriptstyle}
{\nc{\lsim}{\mbox{\raisebox{-.6ex}{~$\stackrel{<}{\sim}$~}}}
{\nc{\gsim}{\mbox{\raisebox{-.6ex}{~$\stackrel{>}{\sim}$~}}}
\def\dsl{\partial\!\!\!/}
\def\lameff{\lambda_{\rm eff}}
\def\Re{{\rm Re\,}}
\def\Im{{\rm Im\,}}
\def\ns{\!\!\!\!\!\!\!\!\!\!\!\!\!\!\!\!\!\!\!\!\!\!\!\!\!\!\!\!\!\!\!\!\!\!\!}
\def\NS{\ns\ns\ns\ns}
\def\dsl{\partial\!\!\!/}
\def\Pl{P_{\sss L}}\def\Pr{P_{\sss R}}
\def\VEV#1{\langle #1 \rangle}
\def\sfrac#1#2{{\textstyle\frac{#1}{#2}}}
\def\ems{{\epsilon_{\rm \bar{MS}} }}
\def\etat{{\eta_t}}
\def\etab{{\eta_b}}
\def\zetat{{\zeta_t}}
\def\zetab{{\zeta_b}}
\def\ttL{{\tilde t_L}}
\def\ttR{{\tilde t_R}}
\def\tbL{{\tilde b_L}}
\def\tbR{{\tilde b_R}}
\def\ttp{{\tilde t_+}}
\def\ttm{{\tilde t_-}}
\def\ttpm{{\tilde t_\pm}}
\def\tbp{{\tilde b_+}}
\def\tbm{{\tilde b_-}}
\def\tbpm{{\tilde b_\pm}}
\def\tqp{{\tilde q_+}}
\def\tqm{{\tilde q_-}}
\def\tqpm{{\tilde q_\pm}}
\def\tqL{{\tilde q_L}}
\def\tqR{{\tilde q_R}}
\def\sss{{\scriptscriptstyle}}
\def\nn{{\nonumber}}
\def\bar{}


\begin{titlepage}
\pagestyle{empty}
\baselineskip=21pt
\rightline{McGill/97-7}
\rightline{hep-ph/9705201}
\rightline{April 30, 1997}
\vskip .4in

\begin{center} {\large{\bf Supersymmetric Electroweak Phase Transition:\\
            Dimensional Reduction versus Effective Potential}}
\end{center}
\vskip .1in

\begin{center} James M.~Cline

{\it McGill University, Montr\'eal, Qu\'ebec H3A 2T8, Canada}

and

Kimmo Kainulainen

{\it  High Energy Physics Division,\\
P.o.\ Box 9, FIN-00014, University of Helsinki 
}
\end{center}

\centerline{ {\bf Abstract} }
\baselineskip=18pt
\vskip 0.5truecm

We compare two methods of analyzing the finite-temperature electroweak
phase transition in the minimal supersymmetric standard model:  the
traditional effective potential (EP) approach, and the more recently
advocated procedure of dimensional reduction (DR).  The latter tries to
avoid the infrared instabilities of the former by matching the full
theory to an effective theory that has been studied on the lattice. We
point out a limitation of DR that caused a large apparent disagreement
with the effective potential results in our previous work.  We also
incorporate wave function renormalization into the EP, which is shown
to decrease the strength of the phase transition.  In the regions of
parameter space where both methods are expected to be valid, they give
similar results, except that the EP is significantly more restrictive
than DR for the range of baryogenesis-allowed values of $\tan\beta$,
$m_h$, the critical temperature, and the up-squark mass parameter
$m_U$.  In contrast, the DR results are consistent with
$2\lsim\tan\beta\lsim 4$, $m_h<80$ GeV, and $m_U$ sufficiently large to
have universality of the squark soft-breaking masses at the GUT scale,
in a small region of parameter space.  We suggest that the differences
between DR and EP are due to higher-order perturbative corrections rather
than infrared effects.

\end{titlepage} \baselineskip=18pt
\textheight8.3in\topmargin-0.0in\oddsidemargin-.0in 

\section{Introduction}

Recently a new method has been proposed and exploited to try to improve 
the reliability of analyzing first-order phase transitions in
finite-temperature gauge theories, called dimensional reduction (DR)
\cite{KLRS2}. In the traditional effective potential (EP) approach, a
problem is posed by the transverse gauge bosons which are very light, 
and thus can lead to infrared instabilities in the effective theory.
Their effects should in principle be treated nonperturbatively to 
reliably compute the strength of the transition. Dimensional reduction
works around this by integrating out all the heavy modes in a
given theory to obtain an effective theory of the problematic light 
modes, which is then studied on the lattice \cite{KLRS1}. This
program was carried out in refs.~\cite{KLRS2,KLRS1} for the standard
model. However, the nonperturbative lattice results of ref.~\cite{KLRS1}
are applicable to other theories as well, since the only requirement is
that the effective theory after integrating out the heavy fields has the
same degrees of freedom as does the standard model, namely the
light Higgs doublet and the transverse gauge bosons.

One of the most interesting applications of DR is to supersymmetric (SUSY)
models, since these have the possibility of producing the baryon asymmetry
of the universe at the electroweak phase transition \cite{HN}.  In
order to preserve the baryons thus created, the phase transition must be
strongly enough first order, meaning in this context that the VEV of the
Higgs fields must be sufficiently large at the critical temperature:
\beq
   v(T_c)/T_c \equiv (v_1^2+v_2^2)^{1/2}/T_c \gsim 1 - 1.5
\label{sphcond1}
\eeq 
Otherwise anomalous baryon-violating interactions mediated by
electroweak sphalerons will be too fast in the broken phase after the
transition, and quickly wash out the baryon asymmetry that was
created. The range of values on the right hand side of (\ref{sphcond1})
comes from an estimate of the uncertainty in the sphaleron rate at
two loops, 
and in dynamical details like the amount of reheating in the phase 
transition \cite{KLRS1}. In the DR approach, condition 
(\ref{sphcond1}) is replaced by the requirement that the ratio of the
Higgs quartic coupling to the squared gauge coupling must be small
enough  in the effective three-dimensional theory,
\beq
    x_c \equiv { \bar\lambda_3/  \bar g_3^2} < 0.026 - 0.044.
\label{sphcond2}
\eeq
The two quantities are related to each other 
by lattice measurements of $v(T_c)/T_c$ in ref.~\cite{KLRS1}; we have
used their results to obtain the fit
\beq
	v(T_c)/T_c \simeq 1.087w^2-2.916w+2.911; \quad
	w\equiv x_c/0.04,
\label{foot1}
\eeq
which is valid in the interval corresponding to (\ref{sphcond1}) and
(\ref{sphcond2}).

References \cite{CK1}-\cite{FL} have studied the phase transition in
the minimal supersymmetric standard model (MSSM) using DR, identifying
the regions in the SUSY parameter space consistent with condition
(\ref{sphcond2}).  One of the purposes of this letter is to explain the
differences between our results \cite{CK1} and the others
\cite{Laine}-\cite{FL}.  In the process we shall elucidate a possible
shortcoming of DR: even moderately heavy squarks can be too light to
reliably integrate out if the superrenormalizable cubic scalar
couplings of the MSSM become too large. 

More generally we are interested in determining the extent to which the
results of dimensional reduction differ from those of the effective
potential (EP) approach \cite{BEQZ}-\cite{DCE}.  We have therefore
undertaken a comparison of the two methods in the MSSM, exploring as
broad a region of parameter space as possible.  We find that they give
results which are in reasonable qualitative agreement.  A further goal
is to characterize which regions of MSSM parameter space are suitable
for electroweak baryogenesis.  We have thus recomputed the
distributions of MSSM parameters, as well as some observables, that are
baryogenesis-compatible.  Where comparison is possible, our results
agree well with some \cite{Laine} of the previous DR studies that have
focused on more specific regions of parameter space, and less well with
others \cite{FL}.

The dimensional reduction procedure employed here is essentially the
same as we used in ref.~\cite{CK1} (hereafter called CK), which
includes the one-loop corrections proportional to the top and bottom
yukawa couplings, $y_{t,b}$.  One must integrate out the third family
quarks and squarks at zero and at finite temperature to find the
effective finite-$T$ lagrangian and to relate its couplings to physical
quantities. The combination $\bar\lambda_3/\bar g_3^2$ in
eq.~(\ref{sphcond2}) can thus be expressed a function of squark masses
and mixing angles, Higgs boson masses and $\tan\beta$, allowing one to
infer from the inequality (\ref{sphcond2}) which are the
baryogenesis-compatible regions in the space of physical parameters.

\section{Dimensional reduction: subleading corrections}

Because $x_c$ needs to be small for purposes of baryogenesis, we are
interested in cases where there are large cancellations between the
tree-level and one-loop contributions to $\bar\lambda_3$.  For
quantitative accuracy it therefore appears important to include some
formally subleading contributions in the finite-$T$ effective
lagrangian, which were omitted in CK. Among the potentially most 
important such terms are those due to thermal loops of gauge and Higgs bosons,
which shift the Higgs boson mass terms \cite{Laine} by an amount
\beq
  \delta {\cal L}_3/T = \left(\frac{1}{4} g^2 T^2 + 
  \frac{3g^2}{16\pi}Tm_{A_0}\right) (|H_1|^2+|H_2|^2),
\label{gTmasses}
\eeq 
where  
  \beq m^2_{A_0} = \frac{5}{2}g^2T^2 + \frac{g^2}{8\pi^2}(3m^2_Q+m_A^2)).
\label{mA0}
\eeq 
These additional terms lower the estimate of the critical
temperature from that found in CK; they alone can increase the
final value of the parameter $x_c$ by $\sim 0.008$. However, there 
is also another, direct contribution to $x_c$ from the same 
particles,
\beq
  \delta x_g =  \frac{g^2\ln 2}{32\pi^2} \cos^2\!2\alpha  - \frac{3
  g^2T}{128\pi m_{A_0}},
\label{directg}
\eeq 
where the angle $\alpha$ defines the direction of the phase transition 
in the $(H_1,H_2)$-plane.  Both in (\ref{gTmasses}) and (\ref{directg}) 
the first term comes from superheavy (nonzero Matsubara frequency) and 
the second from heavy scale (zero Matsubara frequency) integration. The
correction ({\ref{directg}) is $\sim -0.003$, partly cancelling the
effect on $x_c$ from the thermal self-energies in
(\ref{gTmasses}); we argue that this combined effect of $\sim 0.005$ is
the typical scale of dominant $g^2$-corrections.

Taking into account these improvements, our full result for $x_c$ 
can be written as
\beqa
 x_c
 &=& {g^2+g'^2\over 8g^2}\cos^2\!2\alpha
   \nonumber\\
 &+& {3\ln 2\over 4\pi^2}\left({y_t^4\over g^2}\sin^4\!\alpha
         -\frac{g^2+g'^2}{4g^2} y_t^2 \cos 2\alpha\sin^2\!\alpha  \right)
   \nonumber\\
 &-& \frac{y^2_t(g^2+g'^2)}{4g^2}\,\left(\frac{\zeta(3)}{12\pi^4T^2} +
     \frac{T}{4\pi M^3_{D_t}}\right)
     \left(\mu^2 \cos^2\!\alpha - A^2_t \sin^2\!\alpha\right)\cos 2\alpha 
   \nonumber \\
 &+& \left(t\leftrightarrow b;\,\cos\alpha \leftrightarrow \sin\alpha 
 \right),\nn\\
 &+& \delta x^{SH} + \delta x^{H} + \delta x_g 
\label{xc}
\eeqa 
Here $M_{D_t}$ is the sum of the Debye masses left- and right-handed 
top squarks \cite{CK1},
\beq
	M_{D_t} = m_{\tilde t_L} + m_{\tilde t_R},
\eeq
and the thermal corrections are computed assuming the gauginos and
higgsinos are decoupled.
The first three lines in (\ref{xc}) are, respectively,
the contributions from tree-level, leading superheavy scale, and 
the superheavy and heavy scale wave function renormalizations.
The fourth line shows how to obtain the effect of the bottom
sector from that of the top (notice that the interchange of $\sin\alpha$
and $\cos\alpha$ implies $\cos2\alpha\to-\cos2\alpha$).
The last line gives the next-to-leading superheavy scale
shift $\delta x^{SH}$, the much larger term $\delta x^{H}$
from the same Feynman diagrams in the heavy scale integration, and the
gauge boson term from (\ref{directg}).  The additional
function $\delta x^{SH}$ is given by
\beqa
  \delta x^{SH} &=& \frac{1}{g^2}
  \biggl(
    -\frac{3}{2} y_t^4\sin^2\!\alpha
     \sum_{a=Q,U}\left( \sin^2\!\alpha f^{SH}_2(m_a,m_a) - 
     2  S^t_af^{SH}_3(m_Q,m_a,m_U) \right) 
  \nn \\ && \phantom{HH}
    - \frac{3}{2} y_t^4 (  S^t_a)^2 f^{SH}_4(m_Q,m_Q,m_U,m_U)
    - \frac{3}{2} g^2 y_t^2 F^{SH}_\alpha (m_Q,m_U) \; \cos 2\alpha
  \nn \\ 
    && \phantom{Han} + \left( t\leftrightarrow b; \; U 
       \leftrightarrow D; \; \sin \alpha
       \leftrightarrow \cos \alpha \right) \biggr) 
  \nn \\
    && + \frac{g'^2}{4g^2}
    \biggl( y_t^4(F^{SH}_\alpha(m_Q,m_U) - 4F^{SH}_\alpha(m_U,m_Q)) \nn \\
    && \phantom{+l \frac{g}{4}}
       + y_b^4(F^{SH}_\alpha(m_Q,m_D) + 2F^{SH}_\alpha(m_D,m_Q)) \biggr) 
       \cos 2\alpha 
\label{dxi}
\eeqa 
with
$  F^{SH}_\alpha (m_a,m_b) \equiv \sin^2\!\alpha f^{SH}_2(m_a,m_a) - 
     S^t_\alpha f^{SH}_3(m_a,m_a,m_b) $, 
 $   S^t_\alpha \equiv (\mu\cos\alpha+A_t\sin\alpha )^2$. 
The high-temperature expansions of the $n$-point loop integrals
$f^{SH}_n$  (following the notation of \cite{Laine}) are 
\beqa 
  f^{SH}_2(m_a,m_a) &\simeq& -
  \frac{\zeta(3)}{64\pi^4}\frac{m_a^2}{T^2}\nn\\ f^{SH}_3(m_a,m_a,m_b)
  &\simeq& \frac{\zeta(3)}{128\pi^4T^2}  -
  \frac{\zeta(5)}{128\pi^6} \frac{2m_a^2+m_b^2}{T^4} \nn\\
  f^{SH}_4(m_a,m_a,m_b,m_b) &\simeq& \frac{\zeta(5)}{1024\pi^6T^4}
\label{fSH}
\eeqa 
to leading order in $m^2/T^2$.  The contribution from the heavy scale,
$\delta x^{H}$, has the same form as  $\delta x^{SH}$; one
merely replaces the masses $m_{Q,U,D}$ by the corresponding Debye
masses (see ref.\ \cite{CK1}) and the integrals (\ref{fSH}) by 
their heavy scale counterparts, which are exactly given by
\beqa 
  f^{H}_2(m_a,m_a) &=& \frac{T}{8\pi m_a} \nn\\ f^{H}_3(m_a,m_a,m_b)
  &=& \frac{T}{8\pi m_a(m_a+m_b)^2} \nn\\ f^{H}_4(m_a,m_a,m_b,m_b) &=&
  \frac{T}{8\pi m_am_b(m_a+m_b)^3}. 
\label{fi}
\eeqa 
In CK we omitted $\delta x^{HS}$ and the corresponding
corrections to the wave function renormalization, since they are 
formally subleading. The combined effect of all the new terms
we consider here is an increase in $x_c$ of varying size, but never 
larger than $0.015$.

\section{Constraint on superrenormalizable couplings} 

Despite the significance of the subleading contributions which
we now take into account, the most important difference between this
work and CK is that we  have imposed a consistency condition on the
size of the $\mu$ and $A_{t,b}$ parameters of the MSSM, the necessity of
which was not recognized previously.  These couplings are distinguished
by the fact that they appear in superrenormalizable interactions among
the squark and Higgs bosons in the MSSM Lagrangian:
\beq
  {\cal L}_{\rm\sss MSSM} = 
  y_t\,\tilde t_L^* ( \mu  H_1 + A_t H_2) \tilde t_R
 + y_b\,\tilde b_L^* ( \mu  H_2 + A_b H_1) \tilde b_R
      + \cdots
\label{superr} 
\eeq  
Because of their dimensionality, one must be especially vigilant 
against the breakdown of the heavy scale perturbative expansion due to
large infrared contributions to the three-dimensional finite-$T$ 
effective Lagrangian (where the effective quartic coupling has dimensions 
of $T$), of order 
\beq
  \delta\bar\lambda_3\sim 
	{y^2_t T\over 16\pi M^4_{D_t}}|\mu H_1 + A_t H_2|^4, 
\label{muAt}
\eeq 
coming from diagrams like figure 1.  In the present study we have
imposed an upper limit on the ratio of $\mu$ and $A_{t,b}$ to
$M_{D_{t,b}}$ to safeguard against such a breakdown.  
Precisely, we required that 
\beq
   \frac{|\mu|}{M_{D_{t,b}}},\,\frac{|A_{t,b}|}{M_{D_{t,b}}} < 1;
	\ \hbox{and}\ |\mu|,|A_{t,b}| < 500\ \hbox{GeV}.
\label{safecond1}
\eeq

\bigskip
\begin{figure}
\epsfbox{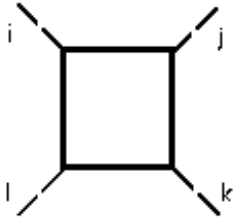} 
\baselineskip=16pt
\caption {A contribution to the effective quartic coupling
of Higgs bosons involving the interactions of eq.~(\ref{superr}), and
generating the term (\ref{muAt}) in the effective Lagrangian. At finite
temperature it is infrared divergent if the masses of the squarks in the
loop vanish.}
\vskip 0.25in 
\end{figure}
\baselineskip=18pt

The precise values chosen for the bounds (\ref{safecond1}) are somewhat
arbitrary.  However the region of MSSM parameter space that satisfies
the sphaleron bound but violates (\ref{safecond1}) appears to be an
isolated ``island'' which, once excluded by (\ref{safecond1}), is not
further reduced by making (\ref{safecond1}) more stringent.  These new
consistency conditions make a dramatic change in the regions of MSSM
parameter space which are most likely to give a strong phase
transition, and are the main reason for the differences between the
present results and those in CK.  There our baryogenesis-allowed points
were dominated by surprisingly high critical temperatures near 300
GeV.  We have subsequently investigated such large-$T_c$ points using
the EP and found that for them, $v/T$ falls below 1 very rapidly with a
slight increase in $T$.  Since the present estimate of $T_c$ (the
temperature where the second derivative of the potential at zero field
has vanishing determinant) is already an underestimate of the true
$T_c$ (where there are degenerate minima of the potential), such
behavior means that the transition is quite unlikely to be strongly
enough first order for baryogenesis.  We find a clear separation
between these spurious solutions to the sphaleron bound, and the
acceptable ones with $T_c$ near 100 GeV, on which we focus henceforth.

The conditions (\ref{safecond1}) also constitute an important difference
between DR and EP. In EP the constraints (\ref{safecond1}) are not
needed, unless one made an expansion of the field-dependent mass
eigenvalues of the squarks in powers of $A_t$ and $\mu$.  But since EP
uses the exact field-dependent tree-level squark masses, it resums
what would be an infinite series of nonrenormalizable contributions to
the effective theory.\footnote{See the caveat discussed below
about resumming the corresponding thermal corrections to the squark
mass matrices}  In DR on the other
hand, one must match the parameters of the effective theory onto the
renormalizable Lagrangian that has been studied on the lattice.  We
therefore have no choice but to truncate the effective theory at the
order of the renormalizable operators. To improve upon this, it will be
necessary to simulate the full MSSM on the lattice, but in the
meantime EP is the only method which has a hope of reliably
determining the strength of the phase transition for values of $\mu$
and $A_t$ where (\ref{safecond1}) does not hold.

\section{The effective potential}

To compare DR to the effective potential approach, we constructed
the EP at the one-loop, ring-improved level, including contributions of
the virtual standard model particles plus the top and bottom quarks
and squarks, and the Higgs bosons.\footnote{The Higgs bosons require
special treatment in the EP because some of them can have negative $m^2$
for small values of $H_i$.  We dealt with this by setting the contributions
to the EP to zero whenever $m^2 < 0$.}  This extends the work described in
ref.~\cite{Brig} where the bottom sector and Higgs bosons were omitted.  
We find that the inclusion of the bottom squarks shifts the critical
temperature significantly ($10-20$\%), but has a small effect on $v/T$,
for experimentally allowed values of the lightest Higgs boson 
mass.\footnote{We explored very large values of $\tan\beta\sim 45$ where the
bottom quark Yukawa coupling would be relevant, but found no cases with
$v/T_c>1$ except those suffering from the same problem as the large $T_c$
points we rejected in connection with (\ref{safecond1}).}
After successfully reproducing the results of ref.~\cite{Brig}, we
further improved the EP by supplementing it with the same wave function
renormalization as we computed in the DR approach.  Usually wave
function renormalization is ignored in the  EP, and the one-loop, 
ring-improved effective lagrangian for the Higgs fields is taken to be
\beqa
  {\cal L}_{\rm eff} &=& \sum_i |D_\mu H_i|^2 - V_{\rm tree}(H_i)
  + \frac{1}{64\pi^2} 
  {\rm Str}\, M^4(H_i) \left(\ln {M^2(H_i)\over Q^2}-\frac32
  \right)\nonumber\\
	&+&{\rm Str}\, T\! \int{ d^3p\over(2\pi)^3}\ln\left(1\pm e^{-\sqrt{
	p^2 + M^2(H_i)}}\right) - \frac{T}{12\pi}{\rm Tr} (M_D^3(H_i,T)
	- M^3(H_i)).
\eeqa
Here Str denotes the supertrace, $\pm$ is $+(-)$ for fermions (bosons),
Tr is the trace over bosons only, $m_D$ is the thermally corrected
Debye mass, and $Q$ is the renormalization
scale.   However it is more accurate to also include the renormalization of
the kinetic term so that
$\sum_i |D_\mu H_i|^2$ becomes $\sum_{ij}Z_{ij} (D_\mu H_i)^\dagger D_\mu
H_j$.  After rescaling the fields to canonical form and ignoring  effects
of two-loop order, this amounts to making the replacement
\beq
  V_{\rm tree}(H_i)\to V_{\rm tree}(\sum_j Z^{-1}_{ij}H_j).
\eeq 
One reason for making this improvement is to try to minimize any
possible sources of discrepancies between DR and EP.  We find that
including wave function renormalization reduces the strength of the phase
transition noticeably, though not drastically.  The comparison is shown in
figure 2, where we have plotted contours of $v(T_c)/T_c$ in the plane of
$\tan\beta$ and $m_A$ (the pseudoscalar Higgs boson mass), for
$m_t = 170$ GeV, $m_Q = 280$ GeV, $m_U = 0$, $A_t = \mu=0$.

To include the effects of the Higgs bosons in the EP, we used the following
for the field-dependent parts of the mass matrices for the CP-even, CP-odd
and charged bosons, respectively:
\beqa
M^2_H(H_i) &=& M^2_H(0) + \frac{T^2}{4}\left(\begin{array}{cc}3y_b^2+g^2
 & 0\\
0 & 3y_t^2+g^2 \end{array}\right) + \frac{g^2}{4}\left\{
\left(\begin{array}{cc}3H_1^2-H_2^2 & -2H_1H_2\\
-2H_1H_2 & 3H_2^2-H_1^2 \end{array}\right),\right. \nonumber\\
&& \left. \left(\begin{array}{cc}H_1^2-H_2^2 & 0\\
0 & H_2^2-H_1^2 \end{array}\right),\quad \hbox{or}\quad 
\left(\begin{array}{cc}H_1^2+H_2^2 & 4H_1H_2\\
4H_1H_2 & H_2^2+H_1^2 \end{array}\right) \right\}.
\eeqa

\begin{figure}
\epsfbox{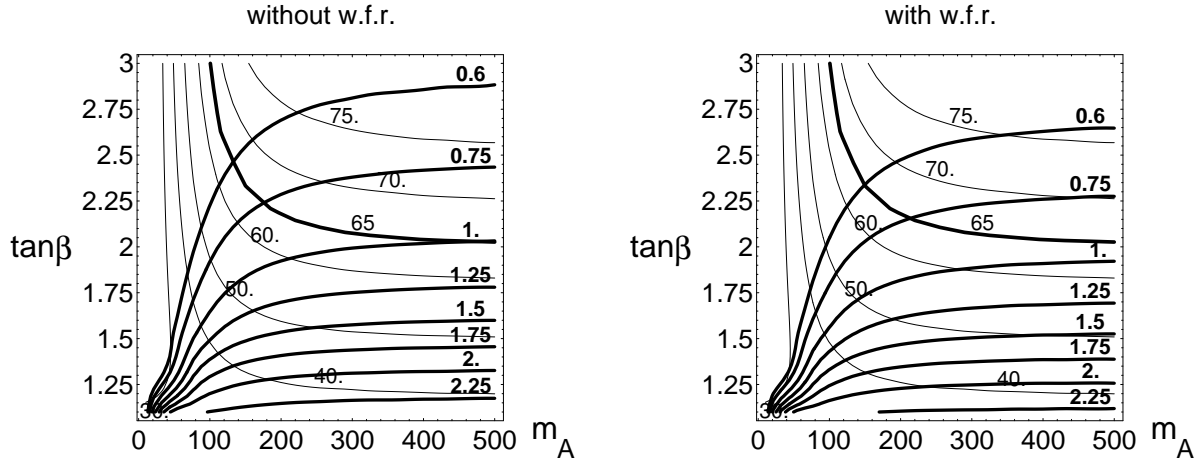} 
\baselineskip=16pt
\caption{Comparison of the usual effective potential
results for the phase transition in the MSSM, to the EP supplemented by
wave function renormalization.  Heavy convex lines: contours of constant
$v/T$ at the critical temperature, in the $\tan\beta$-$m_A$ plane.  Heavy
concave line: contour for $m_h = 65$ GeV.  Light concave lines: contours
for other values of $m_h$.  Fig. 2a is without wave function
renormalization, and 2b is with w.f.r.  The values of the other MSSM
parameters are $m_t = 170$ GeV, $m_Q = 280$ GeV, $m_U = 0$,
$A_t = \mu=0$.} 
\vskip 0.25in 
\end{figure}
\baselineskip=18pt

\section{The high-temperature expansion}

To obtain the above results for the EP, we used the exact expressions
for the $T$-dependent part, without expanding in masses over
temperature.  In contrast, DR is explicitly a high temperature limit,
requiring us to impose the important restriction 
\beq
  m_{\tilde q_X}/T< \pi
\label{debmass}
\eeq  
on the parameters considered, where $m_{\tilde q_X}$ is the thermal
(Debye) mass of any of the squarks included in the loop corrections,
$\tilde q_X = \tilde t_{L,R}, \tilde b_{L,R}$ \cite{CK1}.  One might
therefore naively conclude that EP is valid for a larger range of
parameters than is DR.  However we wish to argue that one should also
make the same restriction (\ref{debmass}) to obtain reliable results
from the EP, because the ring resummation assumes that all the modes in
the ``soft'' loop (heavy scale) can be taken approximately massless,
compared to the modes in the ``hard'' loops (superheavy scale).  Thus the
ring-corrected EP is strictly speaking, and somewhat contrary to 
common wisdom, valid only in high-$T$ limit.  Since the EP has the
qualitatively correct decoupling limit, when one employs the 
nonexpanded integral expressions for loop-corrections however, one 
might hope that the infinite resummation retains some of the qualitative 
physical features of the exact theory, even outside its region of 
strict validity.  Nevertheless, in what follows we will impose 
(\ref{debmass}) equally on DR and EP.

At the same time however, we cannot allow any of the particles we
are integrating out to become arbitrarily light compared to $T$,
because this would give rise to infrared divergences, as discussed above, 
and  necessitate a new lattice study with the light fields included in the
low-energy lagrangian.  Thus we imposed the further constraint 
\beq
  \frac{T}{m_{\tilde q_X}} < 1; \qquad \tilde q_X = \tilde t_{L,R},\
  \tilde b_{L,R}.
\label{safecond2}
\eeq
This ratio appears as an explicit expansion parameter when integrating
out the heavy-scale degrees of freedom; it is the exact 
analog of the expansion parameter $gT/M_W$ in the gauge sector, 
which at small $M_W \sim g\phi$ induced us to use DR rather than EP
in the first place.  We found that EP is quite sensitive to the exact
value taken as the upper limit in (\ref{safecond2}): using a value of 1
removes 80\% of otherwise acceptable parameter sets, whereas using 1.2
leaves essentially all of them.  In the present work we imposed 
(\ref{safecond2}) only on DR, not on EP.

\section {Monte Carlo search of the parameter space}

We undertook a Monte Carlo sweep of the MSSM using both methods, DR and
EP, to find those parameters allowed by the baryogenesis requirement
(\ref{sphcond1}) or (\ref{sphcond2})
\beq 
  x_c < 0.044 \qquad \Leftrightarrow \qquad v(T_c)/T_c > 1.0,
\label{actbound}
\eeq 
and also the
bound on the $\rho$ parameter, which we took to be $\Delta\rho < 0.011$
as the contribution from the third generation quarks and squarks. 
It can be seen below that the scarcity of solutions to (\ref{actbound})
depends strongly on small changes ($\sim 0.001$) in the value of this 
upper bound, so that future 
improvements on the $\rho$ parameter constraint might
severely limit the possibilities for electroweak baryogenesis in the MSSM.
As in CK, the randomly-varied independent parameters  were
$\tan\beta$, $m_t$, the pseudoscalar mass $m_A$, the $\mu$ parameter, 
and the soft
SUSY-breaking parameters $m_Q$, $m_U$, $m_D$, $A_t$ and $A_b$.  The
derived quantities are the physical masses of the squarks and lightest
Higgs boson, the critical temperature $T_c$, $\Delta\rho$, and $v/T$ or
$x_c$.  To compare results, we have converted $x_c$ into the equivalent
$v/T$ value using eq.~(\ref{foot1}).  To determine $v/T$ with the EP,
we numerically solved for the global minimum of the potential, but
found that for the parameters of interest this was never more than 1\%
from the value obtained by minimizing the one-dimensional slice of the
potential in the symmetry-breaking direction determined at the origin.

The resulting distributions of parameters are shown in figure 3, where
one sees reasonable qualitative agreement between the two methods.
However there are several noticeable differences.  The maximum allowed
up-squark mass parameter\footnote{ We excluded negative values of
$m_U^2$ in order to avoid color-breaking minima, although it is
possible to be less restrictive \cite{CQW}. However we did include the
empirical constraint $A_t^2+3\mu^2 < 7.5(m^2_{\tilde t_+}+m^2_{\tilde
t_-})$ from reference \cite{KLS}, since color-breaking minima can occur
even when $m_U^2$ is positive.} is 160 GeV in DR and only 90 GeV in the
EP.  The largest allowed values of $m_{h^0}$, the mass of the lightest
Higgs boson, are 84 GeV and 70 GeV respectively.  In EP we find a
somewhat lower critical temperature (99 GeV versus 92 GeV), and the
distribution of $T_c$ values is also much narrower there than in DR.
The narrowness of the $\tan\beta$ distribution in EP relative to DR has
the anticipated correlation with that of $m_{h^0}$ since the latter
increases in the MSSM with larger $\tan\beta$.  Finally one sees also
that DR has a broader distribution of $v/T$ than does
EP.  The fact that DR has less difficulty than EP to satisfy the
sphaleron constraint is evident in all the parameter distributions
where a difference is discernible.

As for why DR and EP do not agree exactly, we note that the
perturbative expansions are somewhat different in the two approaches.
In particular the ring-improvement in EP is effectively a nonanalytic
resummation with regard to the scalar field dependent terms.  The DR
counterpart of this is the heavy scale integration with the SH-scale
corrected particle (Debye) masses, which is only done to the order of
renormalizable terms in the lagrangian.  Therefore the two methods
differ by contributions that are of two-loop order, and also by
nonrenormalizable terms; we believe these are the source of the
discrepancies.  It is already known that in EP the two-loop 
contributions may considerably strengthen  the
order of the transition \cite{Esp,DCE}, extending the allowed values in
$\tan\beta$ up to $\sim 4$, just as we are now finding in DR but at only
one loop. There may be significant higher-order corrections to DR as well.
Although we imposed the constraint (\ref{safecond2}), to ensure the
overall convergence of the heavy-scale perturbtion expansion, the
two-loop correction may be sizeable when $m_U$ is small.  Indeed one may
roughly estimate that $\delta x_{\rm c,2-loop} \sim (y_t^2/4\pi
)^2(T/m_{\tilde t_L})^2\lsim 0.01$.

\bigskip
\epsfbox{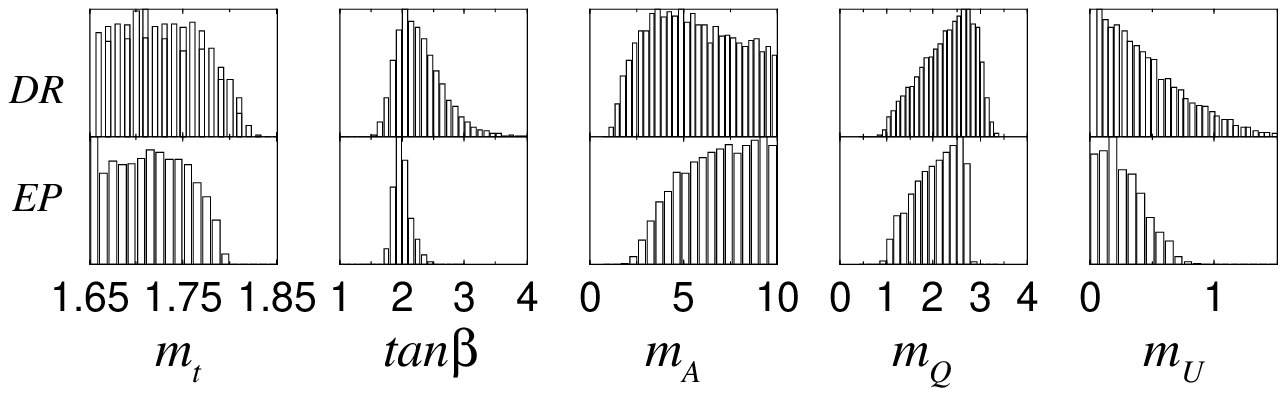}
\epsfbox{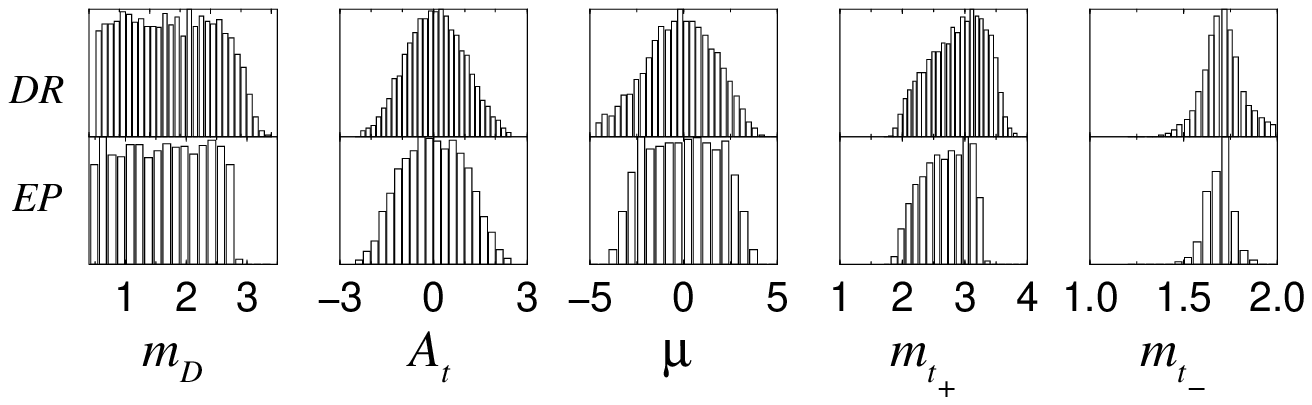}
\epsfbox{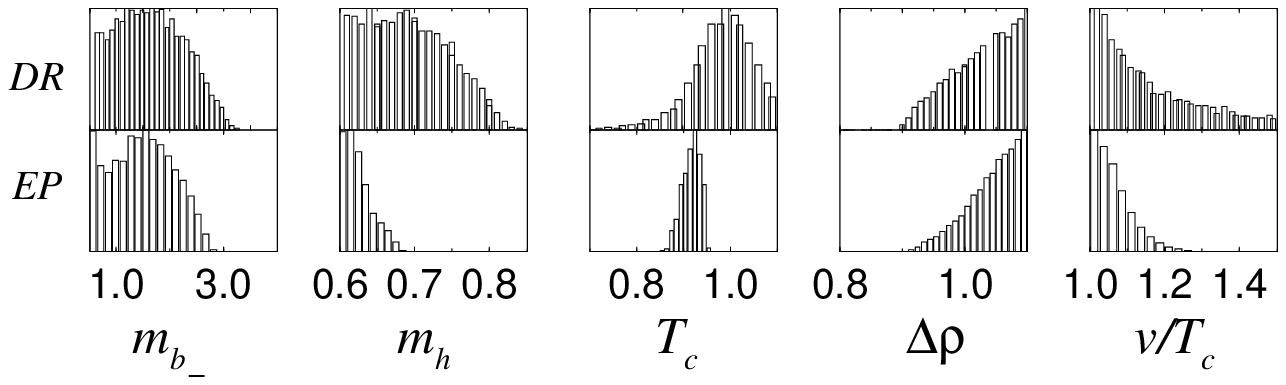}
\baselineskip=16pt
\vspace{-0.25in}
\begin{figure}[h]
\caption{Distributions of MSSM parameters satisfying the
baryogenesis constraint in the dimensional reduction approach (DR) and
in the effective potential approach (EP). Masses are in units of 100
GeV.}
\end{figure}
\baselineskip=18pt

\section {How small must $m_U$ be?}

One might wonder how much fine tuning of parameters is needed to
make the phase transition strongly first order.  One simple assumption is
that all the soft-breaking masses are equal at the GUT scale,
$m_Q=m_U=m_D$, and therefore only differ from each other at the weak scale
by logarithmic corrections.  Using the renormalization group equations to
run the soft-breaking masses down from their universal GUT-scale value
\cite{LT}, and keeping only the corrections due to the large top quark
Yukawa coupling, one can estimate that at the weak scale
\beq
  {m_U\over m_Q} = 
  \left({M_{\sss\rm GUT}\over M_{\rm weak}}\right)^{-y_t^2/16\pi^2}
  \sim 0.7 - 0.8,
\label{massratio}
\eeq 
using $y^2_t = 1.25$ (corresponding to $\tan\beta=2$) and $M_{\sss\rm
GUT} =10^{14}-10^{16}$ GeV.

It is encouraging that this goes in the right direction to be
consistent with electroweak baryogenesis, since the first order phase
transition requires small values of $m_U$.  However the values of
${m_U/m_Q}$ derived from the EP tend to be smaller than this:  the
maximum is 0.8, and values greater than 0.4 are unlikely.  In DR,
${m_U/m_Q}$ has a somewhat less restricted range, going up to 1.6.
Although values as large as in (\ref{massratio}) are relatively
unlikely, at least from the point of view of the parameter space for
the MSSM unconstrained by GUT relations, they are nevertheless
possible.  The distributions for ${m_U/m_Q}$ in DR and EP are shown in
figure 4 (the full extent of the tails of these distributions is not
shown).  Out of 12,500 accepted points in the DR Monte Carlo, about 200
satisfy (\ref{massratio}).  Among these points, the two top squarks are
typically split by less than 80 GeV with $190 < m_{\tilde t_+} < 280$
GeV , the pseudoscalar Higgs boson is heavier than  $m_{A^0} > 180$
GeV, and the right-handed top squark mass parameter lies in the range
$60$ GeV $< m_U < 150$ GeV.  For these points $\mu$ and $A_t$ can be
several hundred GeV but they conspire to give a small value of $\mu +
A_t\tan\beta$ so that the top squark splitting is small. This region
can have a Higgs boson as heavy as $m_h = 79$ GeV.

\bigskip\bigskip
\begin{figure}
\epsfysize=1.5in
\epsfbox{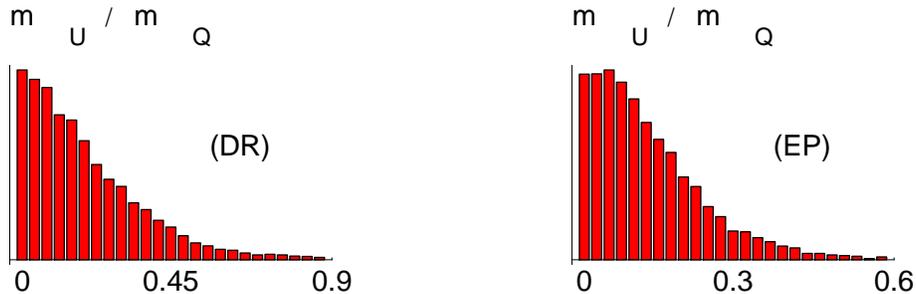} 
\baselineskip=16pt
\caption{Distributions for ${m_U/m_Q}$  in DR and EP.}
\vskip 0.25in 
\end{figure}
\baselineskip=18pt

\section{Correlations between parameters}

In our previous work (CK), there was a strong correlation between the
allowed values of $\tan\beta$ and $m_A$, including points with
arbitrarily large values of $\tan\beta$.  These points were all
associated with large $T_c$, or equivalently with all SUSY breaking
parameters at the TeV scale, as discussed above. Constraints
(\ref{safecond1} and \ref{safecond2}) remove all these points from  our
sample and we find the weaker correlation shown in fig.~5a.  The
correlation between large $\tan\beta$ and small $m_U$ in figure 5b
suggests that the failure of EP to allow large $\tan\beta$ may be due
to the effects of having a very small value of $m_U$.

\bigskip 
\begin{figure}
\epsfysize=2.0in
\epsfbox{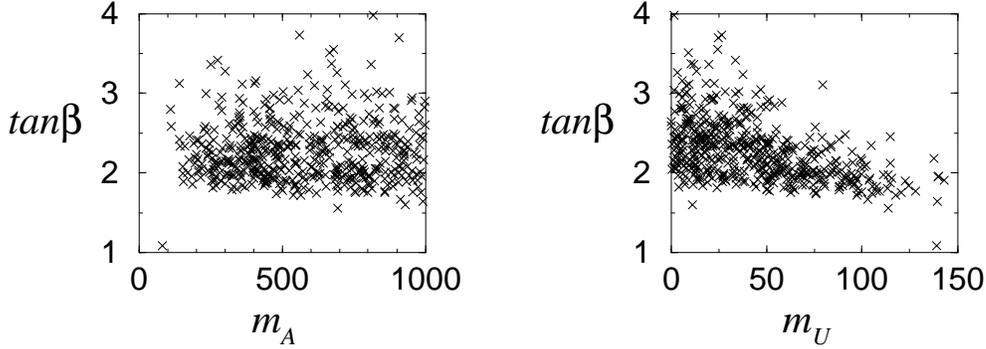} 
\baselineskip=16pt 
\caption{ (a) Baryogenesis-allowed region of MSSM parameter space in DR,
projected on (a) the plane of $m_{A^0}$ versus $\tan\beta$ and (b) in
$m_{U}$ versus $\tan\beta$.} 
\vskip 0.25in
\end{figure}
\baselineskip=18pt

Previous studies of the phase transition in the MSSM have emphasized
that large values of the $\mu$ and $A_t$ parameters have the effect of
weakening the transition.  While this may be true while holding all
other parameters fixed, if they are instead allowed to vary, one can
still find values where the transition is strong even for large $\mu$
and $A_t$.  Despite the restriction (\ref{safecond1}) on $\mu$ and
$A_{t,b}$ that was used to obtain the present results, fig.~6a shows
that a large fraction of the $\mu-A_t$ plane is nevertheless
represented. This is worth emphasizing because in the MSSM the
CP-violating phases which are needed for electroweak baryogenesis are
in precisely these parameters.  Therefore the regions with larger
rather than smaller values of $|\mu|$ or $|A_t|$ are the more interesting
ones.

\bigskip \bigskip
\begin{figure}
\epsfysize=2.0in
\epsfbox{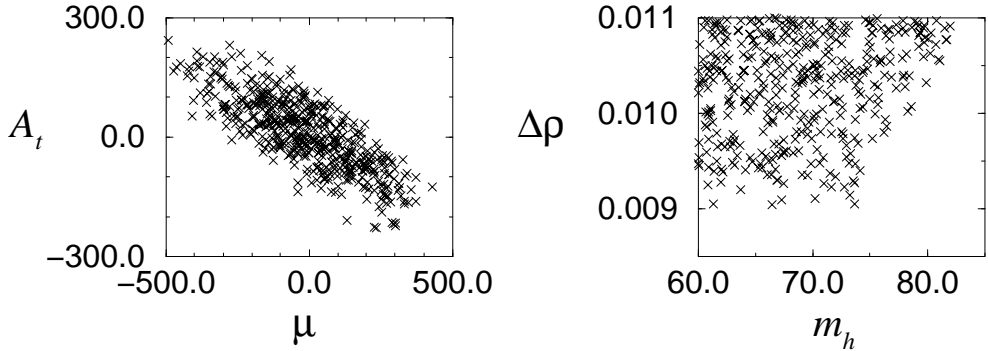} 
\baselineskip=16pt 
\caption{Baryogenesis-allowed region of MSSM parameter space in 
DR, projected on the plane of  (a) $\mu$ versus $A_t$, and subject 
to the conditions (\ref{safecond1}-\ref{safecond2}), and (b) on the
plane of $\Delta\rho$ versus $m_h$.}
\vskip 0.25in
\end{figure}
\baselineskip=18pt

It was mentioned above that the $\rho$ parameter places a stringent
constraint on baryogenesis in the MSSM.  One finds that it is strongly
correlated with various other parameters; in figure 6b we show how in 
DR it appears to be a limiting factor on how large the lightest Higgs
boson mass $m_h$ can be.  The shape of the correlation between $\Delta
\rho$ and $\tan\beta$ looks quite similar, as one might expect from the
tree-level formula $m^2_h = \frac12(m^2_A+m^2_Z - \sqrt{(m^2_A+m^2_Z)^2
-4m^2_A m^2_Z\cos^2\!2\beta})$: it vanishes at $\tan\beta=1$ and increases
for larger $\tan\beta$.

\section{Comparison with other DR results} 

In addition to comparing our DR results with the EP, we have also
compared to other published DR results.  We have analytically checked
that our results agree with those of ref.~\cite{Laine}, up to different
ways of handling the renormalization procedures (which are two-loop
effects), our keeping the finite $g'$ effects in the loops, and some
numerically small additional terms included in ref.\ \cite{Laine}.
Numerically our results agree typically to within less than $0.01$ in
$x_c = \bar\lambda_3/\bar g_3^2$.

More recently, ref.~\cite{FL} gave results which indicated larger
values of the critical ratio $x_c$ than did ours, hence a weaker phase
transition, for given MSSM parameters.  The authors of ref.~\cite{FL}
identified the following possible source for the discrepancy: the
direction of symmetry breaking in the plane of the two Higgs fields is
very sensitive to the critical temperature $T_c$ for small pseudoscalar
mass $m_A$; a large change in this mixing angle can on the other hand
cause large variations in $x_c$.  We have confirmed this expectation by
observing that the tree-level value of $\tan\beta$ (used by ref.~\cite{FL}) 
and the one-loop value (used by us) differ significantly for small values 
of $m_A$.   Indeed, we find the region of $m_A < 100$ GeV to be that 
where the results of all three groups agree least
well.  Fortunately this region is not particularly essential for
baryogenesis, nor for our present results.

The above observation does not fully explain our disagreement with
ref.~\cite{FL} in the large $m_A$ region.  In fig.~7 we show $x_c$ as a
function of $m_A$ at $\tan\beta = 1.75$, for comparison with fig.~2 of
\cite{FL}.  For large $m_A$ our asymptotic value of $x_c \cong 0.03$ is
smaller than theirs by $0.02$.  Of this discrepancy, $0.004$ is due to
the different definitions of $\tan\beta$.  We have also studied the
effects of loop contributions of order $g^2$ to the Higgs boson mass
parameters, neglected by us but included by the authors of \cite{FL},
and find that they can give a further increase of $0.005$.  We are
unable to account for the remaining difference of $0.01$.

\bigskip
\begin{figure}
\epsfysize=3in 
\epsfbox{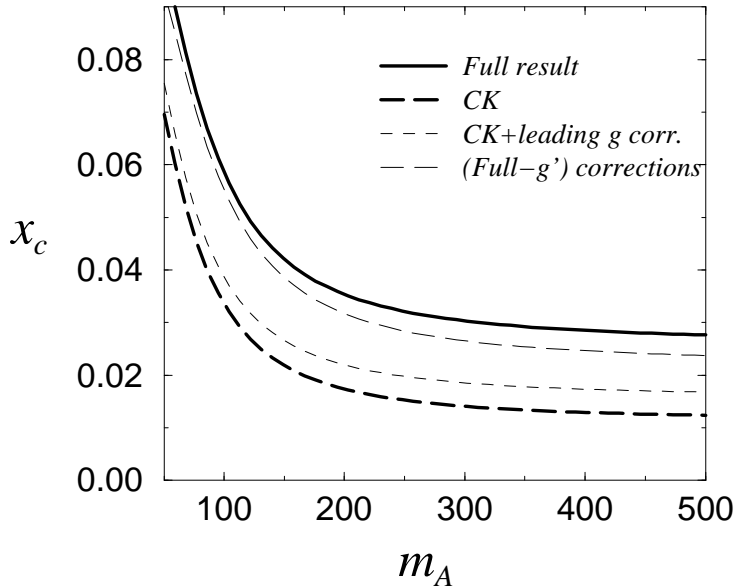} 
\baselineskip=16pt
\caption{ Critical ratio $x_c$ as a function of $m_A$ for
$\tan\beta=1.75$, showing the dependence on corrections of order $g^2$.
Other parameters are $m_Q=265$ Gev and $m_U=100$ GeV, $\mu=A_t=0$. }
\vskip 0.25in 
\end{figure}
\baselineskip=18pt

One should bear in mind that the natural scale of the leading terms
contributing to $x_c$ are of the order of $\sim 0.2$ and the accepted
solutions always correspond to a large degree of cancellation between
these leading contributions.  From this perspective we thus have in
fact a rather good numerical agreement with a relative accuracy of
about 10 per cent with the other published works in the field.  Fig.~7
gives an indication of the sensitivity of $x_c$ to the various
approximations: omitting all $g^2$ corrections, as in CK, putting in
the leading $g^2$ terms but not $g'^2$, and including all the
corrections described in this work.  The difference between
our present work and CK comes dominantly from the neglected squark
sector correction $\delta x^{SH}$, equation (\ref{dxi}).

There are several effects that make it difficult to achieve a better
agreement at this level of approximation.  For example we find that
differences in the definition of heavy scale squark Debye masses
introduce changes of order $0.005$ in $x_c$.  For both DR and EP we
have consistently used the leading order approximations for Debye
masses, whereas one could include also the next-to-leading corrections.
In the absence of the $T=0$ renormalization of the squark sector this
would introduce dependence on the renormalization scale \cite{Laine},
so to get more accurate results the complete renormalization of squark
sector at one loop would be required.  Also, while the other authors
were working in the approximation $g'=0$ for the loop corrections, we
did not; this too makes differences of order $\sim 0.005$ in $x_c$.
Finally, different renormalization procedures cause the definitions of
$m_h$ to differ by terms of two-loop order.  If one uses the physical
Higgs boson mass as an input instead of $\tan\beta$, getting an
accuracy of $0.005$ in $x_c$ would require using the two-loop
computation for $m_h$, since we have checked that a difference of 2-3
GeV in $m_h$ changes the corresponding value of $x_c$ by $0.005$.

\section{Conclusions}

In summary, we have compared the one-loop dimensional reduction and the
effective potential approaches throughout the parameter space of the
MSSM, and found that they give qualitatively similar results for the
strength of the electroweak phase transition, although they differ in
certain quantitative respects, which should be further explored.  We
showed that wave function renormalization has a noticeable weakening
effect on the phase transition when incorporated into the EP.  One
potentially interesting difference is that DR appears to allow a
smaller hierarchy between the soft-supersymmetry-breaking top squark
mass parameters $m_U$ and $m_Q$, which could make SUSY electroweak
baryogenesis compatible with universality at the GUT scale, $m_U=m_Q$.
The top squarks need not have a very large splitting in this case.  We
have also shown that (in either of the methods used) the phase
transition is not necessarily suppressed by large values of $A_t$ and
$\mu$, which is encouraging since they are likely to be the sources of
CP violation if electroweak baryogenesis indeed occurs in the MSSM.

It seems clear that the two-loop corrections are more relevant for
accurate results in the MSSM than for the standard model.  We leave it
for future investigation to see how the differences between DR and EP
are affected when one includes the most important two-loop corrections
in DR, some of which have recently been incorporated into EP in
references \cite{Esp} and \cite{DCE}.  A further shortcoming in the
present state of calculations is that the DR method is no longer
applicable in the MSSM if the $\mu$ or $A_t$ parameters should be
larger than several hundred GeV, or if the top squark becomes too light
during the phase transition.  While the latter could eventually be
accounted for in 3D lattice simulations by including the light squark
field in the effective action, to explore the possibility of very large
$A_t$ or $\mu$ would require a simulation with the full 4D-theory.  Let
us finally point out, in favour of the DR approach, that for reliable
computation of the phase transition dynamics one needs to know not just
the order parameter $v(T)/T$, but also other quanitites like the latent
heat and surface tension, which are not well-approximated by the EP
approach in the standard model \cite{KLRS1}.

\section*{Acknowledgements} 

We thank G.~Farrar and M.~Losada for helpful information about their work, 
and P.~Bamert for discussions about the $\rho$ parameter.

\end{document}